\documentclass{article}
\usepackage{spconf,amsmath,graphicx,caption}

\usepackage{enumitem}
\setlist{nosep, leftmargin=14pt}
\usepackage[top=1in, bottom=1.1in, left=1in, right=1in]{geometry} 

\usepackage{mwe} 
\usepackage{xcolor}

\def\x{{\mathbf x}}
\def\z{{\mathbf z}}

\def\pro{\mathrm{pro}}
\def\retro{\mathrm{retro}}
\def\all{\mathrm{all}}

\newcommand{\bb}[1]{\mathbf{#1}}

\title{Theoretical Bounds on Parallel Imaging Implicit Data Crimes in an MRI Reproducing Kernel Hilbert Space}

\name{Evan Frenklak$^1$, Yamin Arefeen$^{1,2}$, and Jonathan I Tamir$^1$ } 

\address{$^1$Chandra Family Department of Electrical and Computer Engineering, UT Austin, TX, United States \\ $^2$MD Anderson Cancer Center, Houston, TX, United States}
\begin{document}
%
\maketitle

\renewcommand{\thefootnote}{}
   \footnotetext{  
   \copyright{} 2026 IEEE. Personal use of this material is permitted. Permission from IEEE must be obtained for all other uses, in any current or future media, including reprinting/republishing this material for advertising or promotional purposes, creating new collective works, for resale or redistribution to servers or lists, or reuse of any copyrighted component of this work in other works.
   }
\renewcommand{\thefootnote}{\arabic{footnote}}

\begin{abstract}
Magnetic Resonance Imaging (MRI) diagnoses and manages a wide range of diseases, yet long scan times drive high costs and limit accessibility. AI methods have demonstrated substantial potential for reducing scan times, but despite rapid progress, clinical translation of AI often fails. One particular failure mode, referred to as implicit data crimes, is a result of hidden biases introduced when MRI datasets incompletely model the MRI physics of the acquisition. Previous work identified data crimes resulting from algorithmic completion of k-space with parallel imaging and drew on simulation to demonstrate the resulting downstream biases. This work proposes a mathematical framework to re-characterize the problem as one of error reduction during interpolation between sets of evaluation coordinates. We establish a generalized matrix-based definition of the reconstruction error upper bound as a function of the input sampling pattern. Experiments on relevant sampling pattern structures demonstrate the relevance of the framework and suggest future directions for analysis of data crimes. 

\end{abstract}
\section{Introduction}
\label{sec:intro}
Magnetic Resonance Imaging (MRI) is critical for the diagnosis and assessment of wide ranging diseases and anatomy, but long scan time substantially raises costs and precludes equitable access \cite{Hill2021NeurologyOutOfPocket}. Artificial intelligence (AI) methods have substantively demonstrated potential for reducing scan times \cite{heckel2024deep}, but robust clinical translation of AI remains challenging, both in general healthcare \cite{Kelly2019KeyChallengesAI} and accelerated MRI reconstruction. For example, studies have shown that AI reconstruction methods achieve excellent quantitative performance but frequently distort or erase small, critical features \cite{Antun2020InstabilitiesAI} and that the pipeline of training by retrospectively sub-sampling k-space data potentially makes the model unreliable on realistic clinical data \cite{Rajput2024DeathRetrospectiveUndersampling}. 

\textit{Implicit data crimes} have been identified as a critical barrier to the robust clinical deployment of AI for accelerated MRI reconstruction \cite{Shimron2022Implicit}. Open datasets facilitate community wide development of AI, yet many datasets do not contain true raw k-space data, but rather contain processed data that subtly alter the underlying signal statistics. Thus, models trained on such data may perform well in retrospective benchmarks, as processing like zero-padding or JPEG compression reduce data entropy \cite{Shimron2022Implicit}, but will degrade on real clinical data. 
 \begin{figure}[ht]
  \centering
  \centerline{\includegraphics[width=\linewidth]{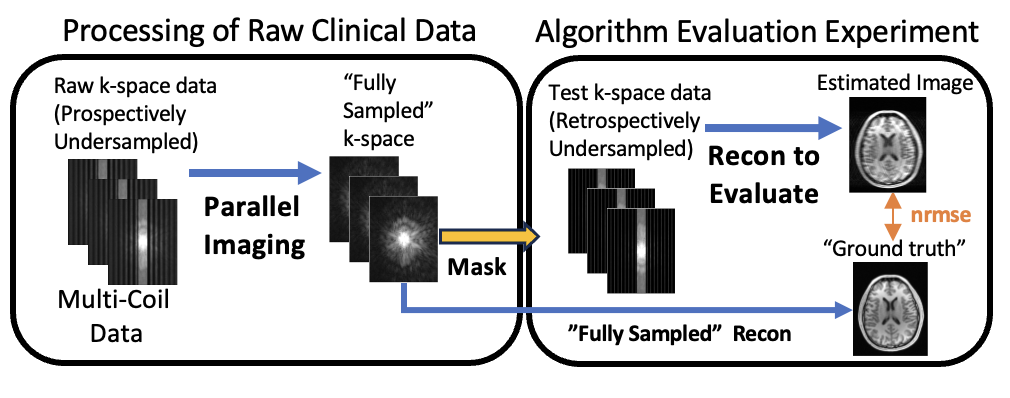}}
\caption{Conceptual framework for the implicit data crime: Prospective sub-sampling occurs in clinical scanner settings. PI fills in missing k-space data before it is presented as a ``fully sampled'' dataset. Retrospectively sub-sampling the algorithmically completed data and comparing it to the PI reconstructed data leads to biased performance evaluation.}
\label{fig:setup1}
\end{figure}

\begin{figure}[ht]
  \centering
  \centerline{\includegraphics[width=1.0\linewidth]{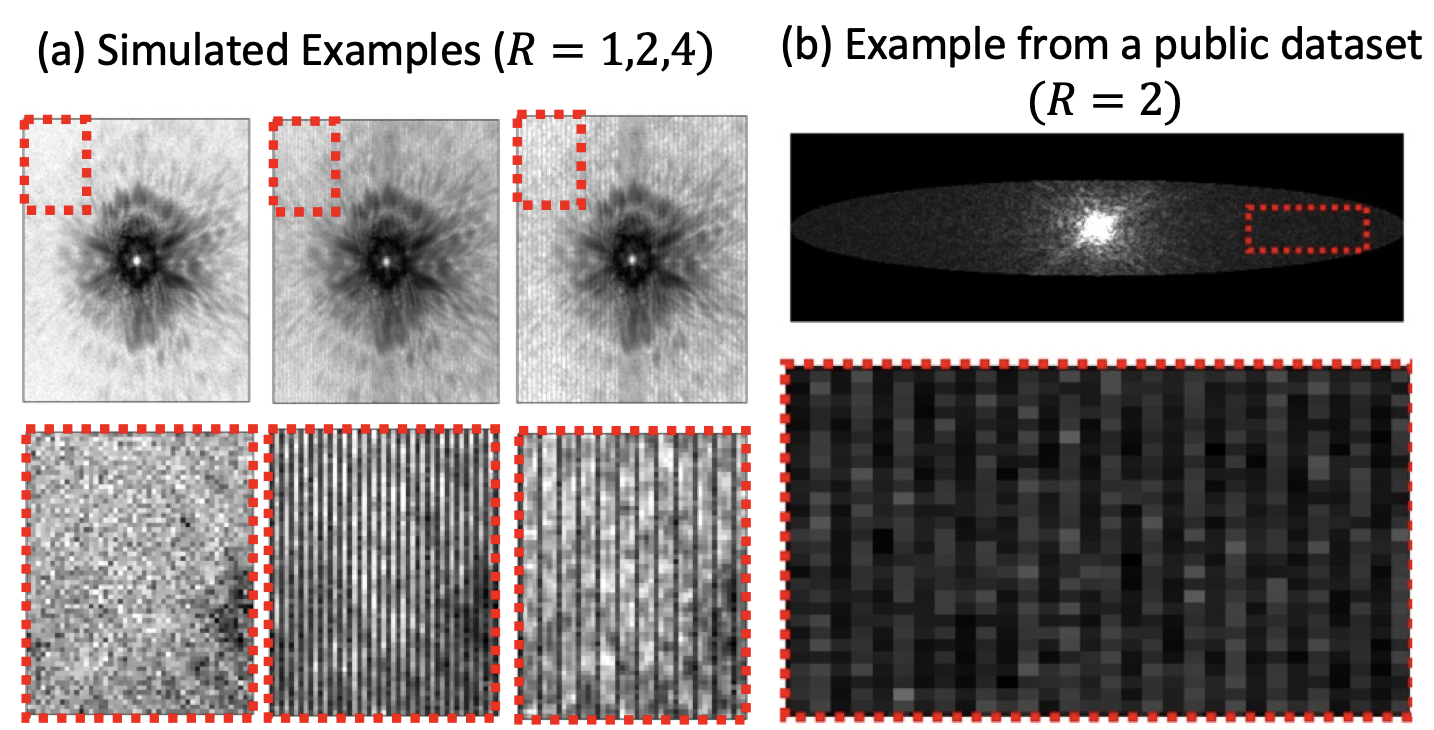}}
\caption{(a) Simulated examples of PI completed k-space and (b) an example k-space completed by PI on a public dataset. Stripes are present in the log-magnitude k-space images because some points have been algorithmically synthesized.}
\label{fig:setup2}
\end{figure}

A recent study identified a new data crime from the observation that most clinical acquisitions measure k-space with sub-sampling due to  the proliferation of multi-coil receive arrays and parallel imaging (PI) \cite{Deshmane2012Parallel}. As a result, curation of data repositories from these clinical acquisitions first involves reconstructing the data with PI. Although these data appear fully-sampled, previous work empirically demonstrated that reconstruction methods evaluated on these algorithmically completed data degrade on realistic, prospective sub-sampled data \cite{frenklak2024parallel}. Figure \ref{fig:setup1} conceptually describes this data crime and Figure \ref{fig:setup2} shows an example of this PI algorithmic completion in a popular public MRI dataset.

In this work, we propose a novel theoretical framework to fully characterize the data crime of evaluating reconstructions on PI-completed data. We extend theory framing PI as linear interpolation in a reproducing-kernel-hilbert-space (RKHS) \cite{Athalye2015ParallelRKHS} by modeling the crime as a two-stage re-sampling process in k-space. From this, we derive reconstruction operators and analytical error bounds corresponding to the two-stage re-sampling process allowing for theoretical characterization and evaluation of the crime. Crucially, our results are data-independent and only rely on the sampling points themselves.

\section{Theory}
\label{sec:theory}

\subsection{Background: RKHS-based MRI Reconstruction}

\begin{figure}[ht]
  \centering
  \centerline{\includegraphics[width=0.95\linewidth]{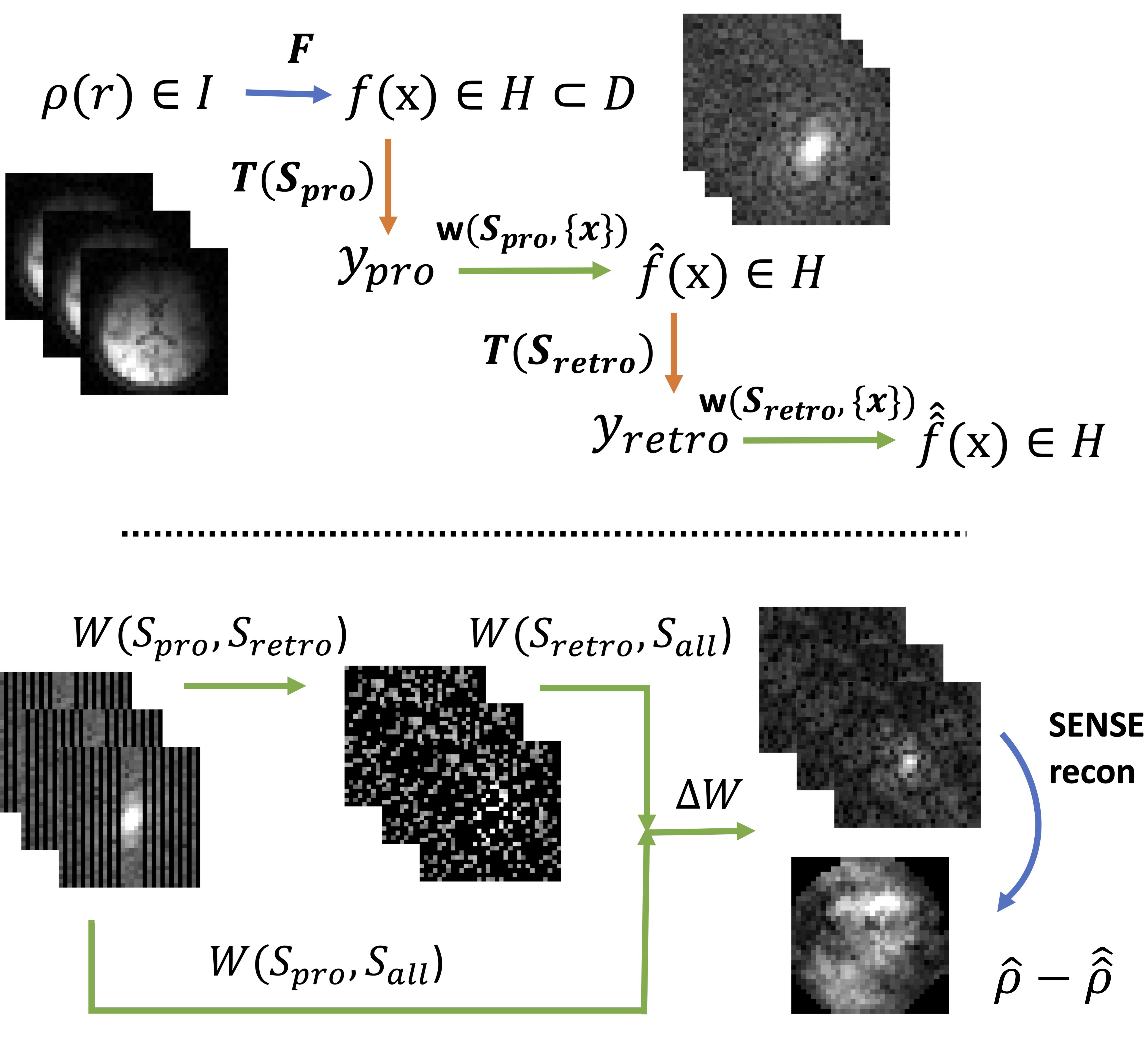}}
\caption{Theoretical and computational framework. (a) $T(S_{\mathrm{pro}})$ and $T(S_{\mathrm{retro}})$ are discrete sampling operators, while elements of the RKHS $H$ are continuous valued, such that $\bb y_{\mathrm{pro}} = \{f_{j_p}(\x_p) | (j_p,\x_p) \in S_{\mathrm{pro}} \}$ and $\bb y_{\mathrm{retro}} = \{\hat{f}_{j_r}(\x_r) | (j_r,\x_r) \in S_{\mathrm{retro}} \}$. (b) The PI data crime is then interpolation between three sets of coordinates. Interpolation weights for the experiment's error are the difference in weights for two reconstruction paths, one passing through $S_{\mathrm{retro}}$ and one skipping it.}
\label{fig:framework}
\end{figure}

Figure~\ref{fig:framework} shows the theoretical and computational framework.
Existing theory \cite{Athalye2015ParallelRKHS} frames PI as linear interpolation in a RKHS, where sampling at k-space point $\mathbf y$ is described by the inner product $f_j(\mathbf{y}) = \langle \epsilon^\mathbf{y}_j,  \rho \rangle$ between the the continuous image, $\rho(\mathbf r)$, and an encoding function 
\begin{align}
\epsilon ^\mathbf{x}_j(\mathbf{r}) =  e^{2 \pi \sqrt{-1} \mathbf{x \cdot r}}  \overline{c_j(\mathbf{r})},
\end{align}
where $c_j$ is the $j^{\mathrm{th}}$ coil sensitivity map and $\mathbf r$ is a spatial coordinate. 
Let $K^\mathbf{x}_j = F\epsilon^\mathbf{x}_j$ be the equivalent k-space sampling function and $M(S_A,S_B)$ be the $|S_A|\times|S_B|$ kernel matrix whose entries 
\begin{align}
M_{a,b} = \langle { K^{\mathbf{x}_a}_{j_a} }, { K ^{\mathbf{x}_b}_{j_b} } \rangle
\end{align}
are inner products between Fourier-domain encoding functions corresponding to samples in either of the ordered sets $S_A$ and $S_B$, where $a$ and $b$ index their respective sets. Each such entry equals the Fourier entry ${F}\{c_{j_a} \overline{c_{j_b}}\}(\mathbf{x}_a-\mathbf{x}_b)$.  Then the RKHS interpolation weights from $S_A$ to $S_B$ are given by 
\begin{align}W(S_A,S_B) = M(S_A,S_A)^\dagger \; M(S_A,S_B),
\end{align}
where $(\cdot)^\dagger$ indicates the pseudoinverse.
RKHS-based reconstruction is optimal in that it has maximum dimension, producing new weights for all data points for each interpolated point, generalizing standard methods like GRAPPA \cite{Deshmane2012Parallel}. 

\subsection{Error Bound Extension to Data-Crime}

Consider a linear retrospective experiment as illustrated in Figure~\ref{fig:framework}. Noisy reference data are prospectively acquired at points $S_{\mathrm{pro}}$ then reconstructed to a fully-sampled grid $S_{\mathrm{all}}$. Test reconstruction does the same after first interpolating to an intermediate pattern $S_{\mathrm{retro}}$. Observe that this chain of reconstructions is equivalent to direct interpolation between the subsets. 

Denoting reconstructions $\hat{f}$ and $\hat{\hat f}$ as in Figure~\ref{fig:framework}, the error reported by the experiment on prospective data vector $\mathbf y_\mathrm{pro}$ can thus be directly interpolated by linear weights
\begin{equation}
\left .(\hat{\hat{f}} - \hat{f}) \right|_{S_{all}}  = \Delta W^\top \mathbf y_\mathrm{pro},
\end{equation}
where 
\begin{equation}
\Delta W = W(S_{\mathrm{pro}},S_{\mathrm{retro}})\; W(S_{\mathrm{retro}},S_{\mathrm{all}}) - W(S_{\mathrm{pro}},S_{\mathrm{all}})
\end{equation}
or some reduced-rank approximations of these matrices.

We next derive a data-independent error bound for retrospective experiment error.
Let the multi-index $\mathbf z_a=(\mathbf{x}_a,j_a)$ contain both the k-space sampling coordinate and the coil index, and let $w(\mathbf z_a,\mathbf z_b)$ be the scalar interpolation entry $W(S_A,S_B)_{a,b}$. We are concerned with the power function of the difference in k-space between the prospective and retrospective reconstructions:

\begin{align}
&\left|(\hat{\hat{f}}-\hat{f} )(\mathbf{z})\right| =
\left|\sum_{\z_r \in S_{\retro}} \hat{f}(\z_r) w(\z_r,\z)  - \hat {f}(\z) \right|
\\
\begin{split}
{}&=\left|\sum_{\z_r} \sum_{\z_p \in S_{\pro}} f(\z_p) w(\z_p,\z_r) w(\z_r,\z)\right . \\
&\quad\quad\quad{}-   \left .\sum_{\z_p \in S_{\pro}} f(\z_p) w(\z_p,\z) \right|
\end{split}\\
&=\left|\sum_{\z_p} \left\langle K^{\z_p} , f \right\rangle \sum_{\z_r} \left[ w(\z_p,\z_r)w(\z_r,\z) - w(\z_p,\z) \right] \right| \\
&= \left|  \sum_{\z_p} \langle K^{\z_p} \Delta w(\z_p,\z) , f \rangle \right| \\
&\leq \left\| f \right\|_H \left\| \sum_{\z_p} K^{\z_p}  \Delta w(\z_p,\z)  \right\|. \label{eqn:pz}
\end{align}
Defining the normed term in (\ref{eqn:pz}) as the power function $p(\z)$, the kernel's reproducing property gives the squared power,
\begin{align}
&p^2(\z) = \left\langle  \sum_{\z_p} K^{\z_p}  \Delta w(\z_p,\z), \sum_{\z_{p'}} K^{\z_{p'}}  \Delta w(\z_{p'},\z) \right\rangle \\
&= \sum_{p \in S_{\pro}} \sum_{p' \in S_{\pro}} 
\left\langle  K^{\z_p} ,  K ^{\z_{p'}}  \right\rangle
\Delta w(\z_p,\z) \overline{\Delta w(\z_p',\z)}.
\end{align}
In matrix form this simplifies to
\begin{equation}
    \mathrm{diag}\left( \Delta W^\top M(S_{\mathrm{\pro}},S_{\mathrm{\pro}}) \overline{\Delta W} \right).
\end{equation}

This result, a point-wise upper bound, characterizes the structure and degree of error in the experiment. PI leverages linear consistency in the signal subspace spanned by the sensitivity maps \cite{haldar_lsi}. This bound describes the relative difficulty of interpolating each point of k-space from the retrospective data versus direct interpolation from the prospective data. Its spatial structure depends on the prospective pattern, the retrospective pattern, and the sensitivity maps.

When $S_\pro=S_\all$, the power function describes the single-stage interpolation error of the retrospective experiment. Such a setting is ``data-crime free'' as all k-space points are Nyquist-samples of the data. On the other hand, when $S_\pro=S_\retro$, the retrospective experiment is trivial, requiring no new interpolation. This case maximizes the ``data-crime'', potentially (aside from possible regularization effects) suggesting an error-free result. 

The norm of columns of the $\Delta W$ matrix gives the noise amplification term $\nu_{j_b}(\mathbf x_b) = \|\Delta w_b(S_A,S_B)\|_2$. It is a vector magnitude of difference in weights for output coordinate $(j_b,\mathbf x_b)$, characterizing the stability of interpolating experiment error there.

\section{Methods}
\label{sec:pagestyle}

Using the BART software toolbox \cite{bart}, a 4-coil, 32$\times$32 phantom was created in k-space so as to avoid an inverse crime \cite{inverse_crime_mri}. We add noise to the data for an overall SNR of 35dB. Regularization set to $10^{-5}$ times the average eigenvalue of $M(S_{\mathrm{all}},S_{\mathrm{all}})$ was added along the diagonal of all inverted matrices for all experiments. We apply the computational framework in Figure~\ref{fig:framework} to compare a few pattern combinations: a fully-sampled prospective grid versus an $R_\mathrm{pro}=2$ 2D uniform sampling pattern for an $R_\mathrm{retro} = 3$ uniform 2D pattern; and an $R_\mathrm{pro}=4$ CAIPIRINHA \cite{breuer_controlled_2005} 3D prospective pattern versus an $R_\mathrm{pro}=4$ random 3D pattern for an $R_\mathrm{retro}=4$ random 3D pattern.

For each experiment, we show the SENSE reconstruction \cite{Deshmane2012Parallel} after the chain of sampling, as well as the corresponding image and k-space errors. We also calculate and display the composite power function after interpolating to a full grid.

\begin{figure*}[t]
  \centering
  \centerline{\includegraphics[width=.7\linewidth]{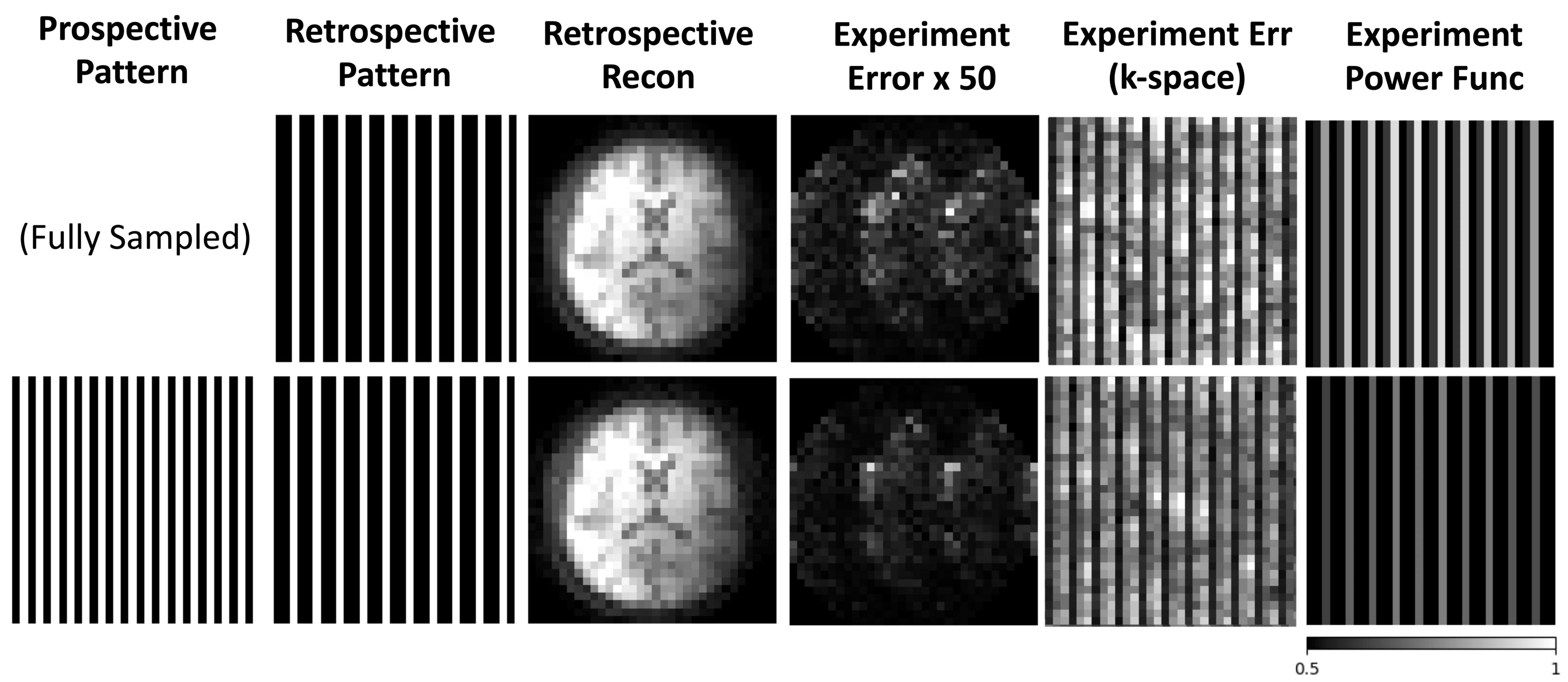}}
\caption{Experiment error and power functions for 2D sampling patterns. Our data-independent power function predicts k-space structure of experiment error, and its magnitude is suppressed due to the introduction of a prospective pattern. }
\label{fig:result}
\end{figure*}
\begin{figure*}[t]
  \centering
  \centerline{\includegraphics[width=.7\linewidth]{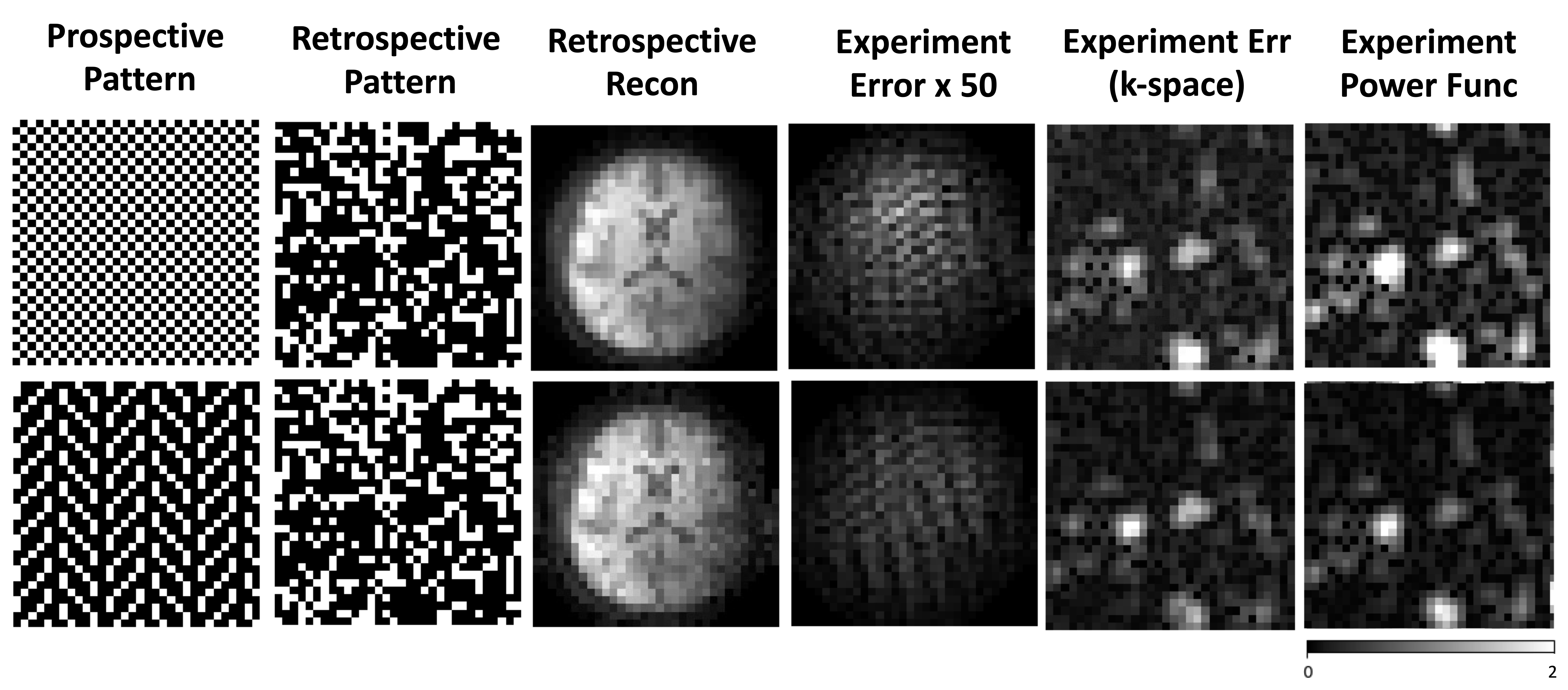}}
\caption{Direct interpolation of retrospective experiment error and power function calculation for 3D sampling patterns.}
\label{fig:result2}
\end{figure*}

\section{Results}
Figure \ref{fig:result} demonstrates the result of introducing a prospective 2D sampling pattern to a retrospective sampling experiment. An acceleration of $R_\pro=2$ is a standard clinical practice, and in fact, parallel imaging reconstructs missing k-space points effectively for a uniformly-spaced pattern. Still, the experiment power function, measuring interpolation difficulty between this pattern and an $R_\retro=3$ uniform pattern along the same dimension, subtly but perceptibly is suppressed in magnitude compared to using a fully-sampled prospective dataset. This follows from the reduced dimension of all functions interpolated from the prospective pattern, that is, a subspace projection induced by that pattern. 

Figure \ref{fig:result2} demonstrates combinations of 3D Cartesian sampling patterns, which might vary across two spatial dimensions. Interpolation to a random 3D pattern from a CAIPIRINHA grid generates a power function that is structured according to interpolation gaps in k-space. As this pattern minimizes gaps between k-space samples in 2D, it leads to a stable prospective reconstruction, such that the power function resembles the structure of the 3D pattern. A final example interpolates from a less stable 3D pattern, outputting a power function with some regions of reduced magnitude, anticipating subtle changes in experimental error. 

\section{Discussion and Conclusion}
Experiments visualize the connection between the power function derived above and the choice of sampling pattern pairs in the PI inverse crime. RKHS weights and their corresponding power functions are expensive to compute, requiring the solution to a linear system whose entries are $\Theta(N^2)$ in each dimension of the input data, and requiring up to $O(N^6)$ time to solve the system for an $N\times N$ image and even longer for 3D data. Matrix-free programs may help make these solutions tractable for full-size problems. For this phantom case, however, RKHS theory successfully predicts the structure of retrospective experiment error for standard MR k-space sampling approaches.  

Composite power functions give rigor to the related data crime phenomenon exhibited in \cite{frenklak2024parallel}. The analysis is general and could be applied to arbitrary (e.g., non-Cartesian) sampling patterns. Upper bounds give useful information about the structure of error in k-space, possibly inspiring future reconstruction regularizers and sampling pattern optimizers, as well as mitigation tools that predict bias due to the PI data crime. While the high computational cost of RKHS interpolaters limits immediate application, these methods relate closely to low-dimensional reconstructors such as GRAPPA and image-domain ones such as SENSE \cite{Deshmane2012Parallel}. Subsequent analysis should bridge linear error analysis to those methods and to non-linear AI-based reconstruction, helping researchers to improve the rigor of their algorithmic pipelines.




\section{Compliance with ethical standards}
\label{sec:ethics}

This is a numerical simulation study for which no ethical approval was required.

\section{Acknowledgements}
\label{sec:acknowledgements}

Evan Frenklak is supported by NSF Award \#2313998, subaward G-1B-022. This work was funded in part by grants
from NSF CAREER CCF-2239687 Award, NSF IFML 2019844, and UT Joint Center for
Computational Oncology Postdoctoral Fellowship.

\bibliographystyle{IEEEbib}
\bibliography{strings,refs}

\end{document}